# Title: Molecular wave-packet dynamics on laser-controlled transition states

**Authors:** Andreas Fischer[1]*, Martin Gärttner[1], Philipp Cörlin[1], Alexander Sperl[1], Michael Schönwald[1], Tomoya Mizuno[1], Giuseppe Sansone[1,2], Arne Senftleben[3], Joachim Ullrich[4], Bernold Feuerstein[1], Thomas Pfeifer[1], and Robert Moshammer[1]

**Affiliations:**

[1]Max-Planck-Institut für Kernphysik, Saupfercheckweg 1, 69117 Heidelberg, Germany.

[2]Dipartimento di Fisica, Politecnico Milano, Piazza Leonardo da Vinci 32, 20133 Milano, Italy

[3]Institut für Physik, Universität Kassel, Heinrich-Plett-Str. 40, 34132 Kassel

[4]Physikalisch-Technische Bundesanstalt, Bundesallee 100, 38116 Braunschweig, Germany.

*Correspondence to: andreas.fischer@mpi-hd.mpg.de

Understanding and controlling the electronic as well as ro-vibrational motion and, thus, the entire chemical dynamics in molecules is the ultimate goal of ultrafast laser and imaging science[1,2]. In photochemistry, laser-induced dissociation has become a valuable tool for modification and control of reaction pathways and kinetics[3–9]. Here, we present a pump-probe study of the dissociation dynamics of $H_2^+$ using ultrashort extreme-ultraviolet (XUV) and near-infrared (IR) laser pulses. The reaction kinematics can be controlled by varying the pump-probe delay. We demonstrate that the nuclear motion through the transition state can be reduced to isolated pairs of initial vibrational states. The dynamics is well reproduced by intuitive semi-classical trajectories on a time-dependent potential curve. From this most fundamental scenario we gain insight in the underlying mechanisms which can be applied as design principles for molecular quantum control, particularly for ultrafast reactions involving protons.

In many control schemes, dissociation is achieved by laser-induced coupling of two electronic states leading to the formation of light-induced potential energy surfaces (LIP)[10,11,9]. It has even been demonstrated that, by using evolutionary algorithms, light-pulses can be shaped to induce a desired reaction[12–16]. The complexity of molecular systems, however, is the main challenge towards comprehensive molecular laser control and the required understanding of the corresponding dynamical mechanisms. This complexity stems from the vast number of degrees of freedom of polyatomic molecules – including electronic transitions, vibrational and possibly rotational states – which have to be controlled. In addition, laser pulses for creating LIPs operate in the strong-field regime, where non-linear dependencies on the laser intensity and pulse duration are encountered. In view of this complexity, the molecular hydrogen ion $H_2^+$ represents the most fundamental testbed of quantum control of ultrafast molecular motion. Moreover, a thorough understanding of this archetypal system is of utmost importance since reactions involving protons are ubiquitous in chemistry and biochemistry like proton transfer and isomerization.

Experimentally, we prepare nuclear wave packets in the potential of the $H_2^+(X\ ^2\Sigma_g^+)$ ground state (|g>), by ionizing molecular hydrogen using XUV attosecond few-pulse trains. Subsequently, we dissociate the molecule using a time-delayed sub-10 fs IR control laser pulse, which creates the LIPs by coupling |g> to the $H_2^+(A\ ^2\Sigma_u^+)$ state (|u>). As the coupling of the two states |g> and |u> is strongly dependent on the field strength of the control pulse, and thus on time, the motion on the LIP through this transition state is no longer energy conserving. In contrast to pump-probe experiments, where such a time-dependence of the potential is unwanted, a control scheme exploits this feature to influence the reactions dynamics. Figure 1 illustrates the dissociative ionization

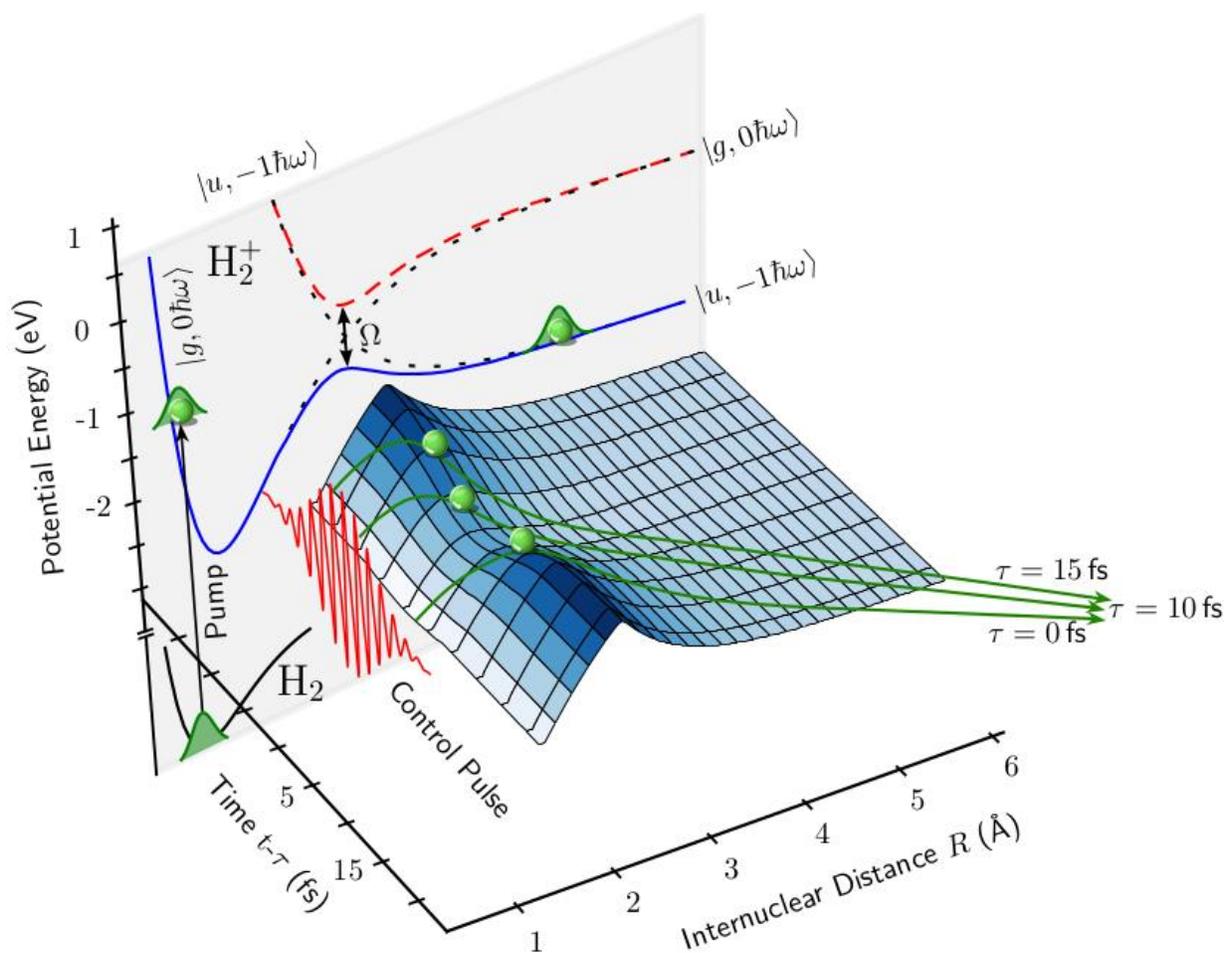

**Fig. 1: Illustration of the experimental pump-control sequence.** Background: The pump pulse ionizes neutral $H_2$, preparing a vibrational wave packet in the $H_2^+$ ground state |g>. The time-delayed control pulse couples |g> to the dissociative |u> state. Blue solid and red dashed curves show the relevant field-dressed potentials (LIPs) (Floquet picture). The opening $\Omega$ of the resulting avoided crossing (i.e., the height of the dissociation barrier) depends on the coupling laser intensity. Foreground: The blue surface shows the time evolution of the lower LIP. The envelope of the control pulse leads to a time-dependent lowering of the dissociation barrier. Depending on the time delay $\tau$, the dissociating fragments either gain or lose energy while traversing the barrier (three example trajectories are shown).

reaction as classical trajectories on the time-dependent LIP. In order to demonstrate the manipulation of the reaction kinematics, we detect the dissociating proton in a kinematically complete experiment[17]. Figure 2 (a) shows the measured H$^+$ counts $C(v_p,\tau)$ as a function of the proton velocity $v_p$ and the time delay $\tau$ between the two pulses, where $\tau<0$ fs corresponds to the case where the IR precedes the XUV pulse.

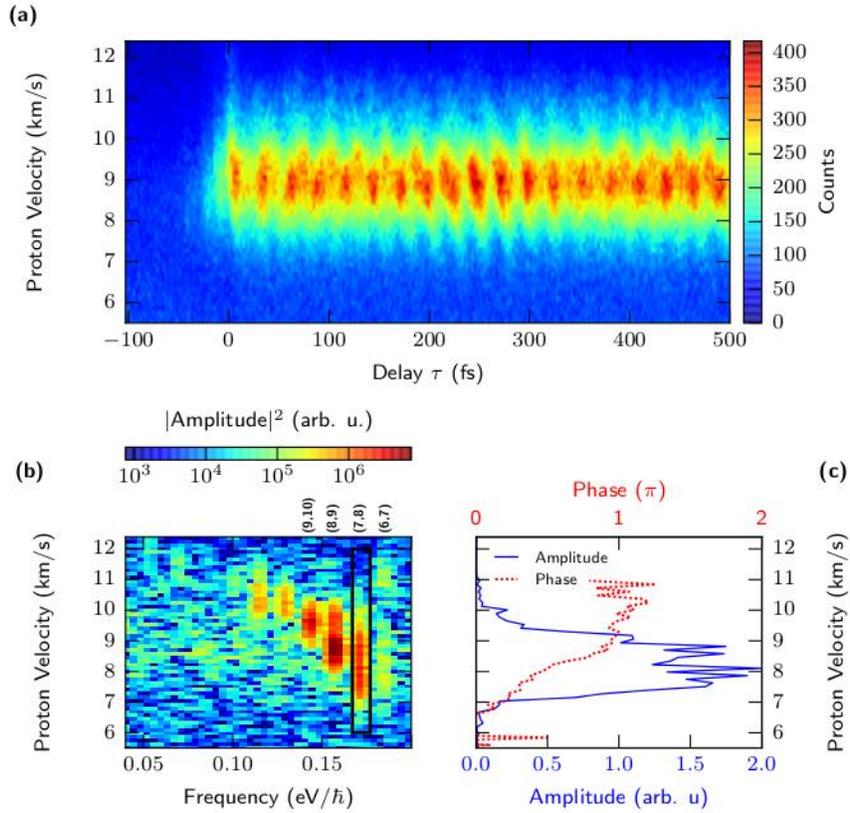

**Fig. 1: Fourier analysis of measured delaytime spectrum.** (a) Measured time-dependent proton spectrum $C(v_p,\tau)$. A pronounced oscillation is visible. (b) The Fourier transform $F(v_p,\omega)$ of $C(v_p,\tau)$ is shown. The lines correspond to the beating of two adjacent vibrational states. (c) The amplitude of the beating of the vibrational states $v=(7,8)$ [framed in (b)] as a function of the proton velocity and the corresponding phase, which encodes the dynamical information of the dissociation process.

A pronounced oscillation with a period of approximately 26 fs is visible. This oscillation reflects the beating between the excited vibrational states. The anharmonicity of the potential-energy curve corresponding to the |g> state causes the observed oscillation to de- and re-phase. This manifests in periodic variations of the oscillation contrast in Fig. 2 (a) (nuclear wave-packet revival). Fig. 1 (b) shows the Fourier transform $F(v_p,\omega)$ of the count distribution $C(v_p,\tau)$. We assign a pair of vibrational states to each line visible in Fig. 2 (b) by comparing the observed beating frequencies to the energy spacings of the vibrational eigenenergies. The measured revival structures are consistent with previous studies of vibrational wave-packet motion, see e.g. Refs. [18–25]. However, the *absolute value* of the Fourier coefficients [solid line in Fig. 2 (c)] of the time-delay spectrum reveals only part of the quantum-mechanical information. Here, we go beyond amplitude-only analysis to directly access the strong-field coupling dynamics of selected states

The key to access the dynamical information is the *phase* of the Fourier coefficients [dotted line in Fig. 2 (c)]. To illustrate the meaning of these phases we select a single line of the Fourier spectrum [e.g. black box in Fig. 2 (b)], corresponding to two vibrational states $v_1$ and $v_2$, and perform an inverse Fourier transform back to the time domain (wavelet analysis). This way, we isolate the part of the time delay spectrum that is oscillating with frequency *$(E_{v2} - E_{v1})/\hbar$*. Physically, this signal corresponds to the result that would be obtained if only two isolated vibrational states were prepared in the first place. We confirmed this by simulating quantum-mechanical wave packets consisting of only isolated pairs of eigenstates (see supplement).

The results of this wavelet analysis are shown in Fig. 3 for two different laser intensities *($I_1=5.1 \cdot 10^{12}$W/cm$^2$ and $I_2=4.5 \cdot 10^{11}$W/cm$^2$)*. For the case of $I_1$ [(a)-(d)] two pairs of vibrational states, namely *v=(7,8)* and *v=(8,9)* are shown. The deeper bound vibrational state pair *v=(7,8)* is no longer dissociated for the case of $I_2$. For increasing intensity, we observe a broadening of the proton velocity distribution. This implies that for high field strengths we enter a non-energy-conserving regime, in which the control of the reaction dynamics becomes possible. In this regime, the final fragment velocity decreases as a function of the time-delay τ with a period corresponding to the quantum-mechanical wave packet.

The mechanism responsible for the (oscillatory) time- and intensity-dependence of the final proton velocity can be understood in terms of a semi-classical model which we discuss in the following. For this, we propagate a classical particle on the time-dependent LIP (blue surface in Fig. 1) by solving Newton's equation. The particle is initially at rest at the left turning point of the binding potential |g>, which energetically corresponds to the mean energies of the modeled vibrational states *$(E_{v1}+E_{v2})/2$*. In the absence of a strong laser-field, the bound nuclei now oscillate in the ground state potential of the molecular ion. In the presence of a laser pulse, the LIP's barrier is lowered. Depending on the relative phase of the particle's oscillation and the strong-field pulse, we obtain trajectories on the LIP with different final state velocities. This is illustrated by the arrows in Fig. 1 and the quantitative results of the simulations are represented by the dotted lines in Fig. 3.

Using the time-dependent LIP picture, we are thus able to explain the mechanism that causes the proton's time-delay dependent velocity. For this, we first note that the LIP in the transition state (coupling region) can be regarded as an "elevator". During the interaction with the coupling laser it moves down for increasing laser intensity and up again for decreasing laser intensity. This up/down motion of the "elevator" happens on the same time scale on which the particles traverse the coupling region. This results in a change of total energy of the system. To clarify this mechanism we consider two cases, which lead to a lowering and to a gain of energy, respectively. First, the particles enter the downward-moving "elevator" and exit it close to its lowest point, and are therefore losing energy. This case is illustrated in Fig. 1 by the arrow labeled *τ=15* fs. In the second case, the particles traverse the coupling region while the "elevator" moves upward and the nuclei gain energy (illustrated by the arrow labeled $\tau=0$ fs). This means that the coupling laser-field acts as a control knob to accelerate or decelerate the nuclei during the molecular dissociation reaction depending on the relative phase between the wave-packet oscillation and the arrival of the coupling pulse. This mechanism is further illustrated in the supplementary movies (Movs. S4-S6).

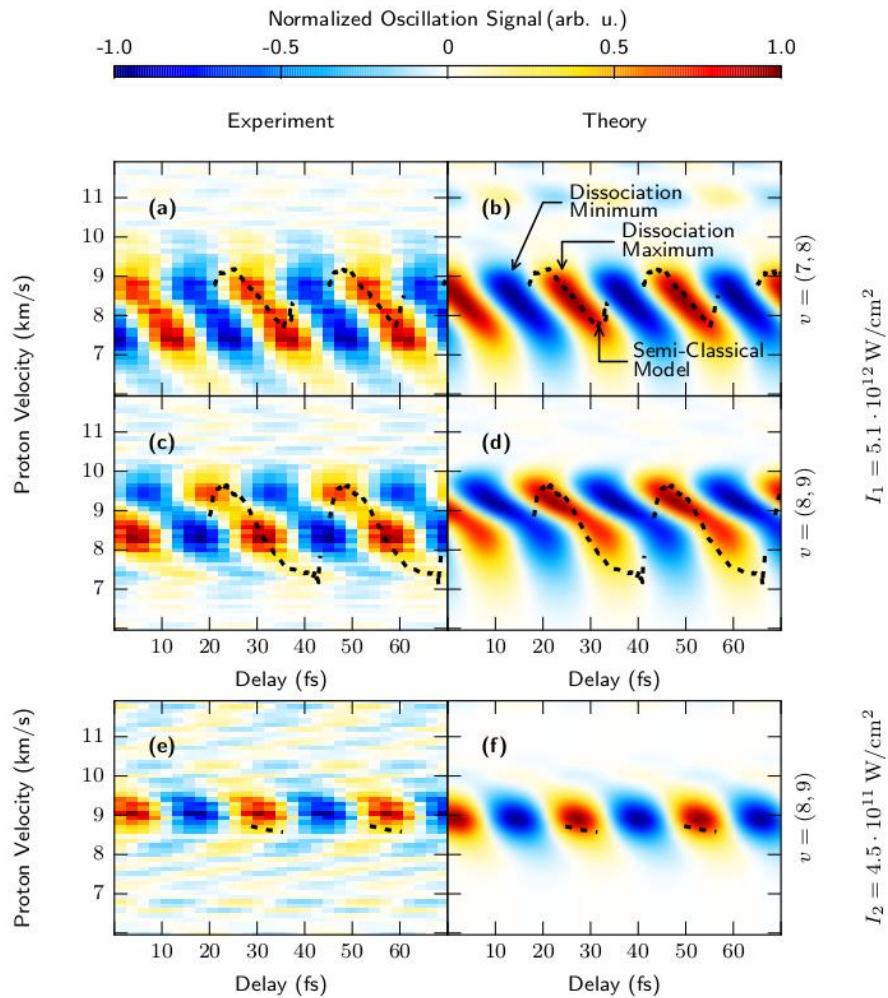

**Fig. 2: Wavelet analysis of Fourier lines.** Analysis of lines corresponding to the pairs of vibrational states $v=(7,8)$ and $v=(8,9)$ for two different laser intensities [(a-d) $I_1=5.1\cdot 10^{12}W/cm^2$ and (e-f) $I_2=4.5\cdot 10^{11}W/cm^2$]. Left column: The experimental data. Right column: The quantum simulation. For intensity $I_2$ the dissociation probability of the states $v=(8,9)$ is vanishingly small, hence we omitted the plot. The signal average along the time axis is subtracted and the spectra are normalized to values between $-1$ and $1$, where $-1$ corresponds to the least dissociation amplitude and $1$ to the highest. The dotted lines show the proton velocity as a function of time delay $\tau$ predicted by the semi-classical simulation (see text).

In conclusion, we have explained the physical mechanism behind strong-field induced velocity control of molecular reaction products. We find an "elevator" mechanism to be responsible for the control of the reaction through a laser-induced transition state. It is surprising that a semi-classical model is capable of describing and predicting the velocity distribution of the dissociation connected to a wave packet consisting of only two vibrational states. The fact that the wave-packet dynamics for light molecules such as $H_2^+$, where quantum effects are expected to be the most significant, can be understood in terms of semi-classical modeling indicates a broad applicability of such processes in quantum control. Especially for larger molecules, where a complete quantum-mechanical modeling is difficult if not impossible, the validity of semi-classical mechanisms proves essential to reduce the complexity of the system's description. The control over internuclear velocities by the discovered "elevator" mechanism promises an efficient approach to specifically address different intermediate and final states even in non-dissociating scenarios. Applied to larger molecules, this mechanism thus provides a rational design principle for strong laser pulses acting as reagents for future applications in laser chemistry [see e.g. Ref. [26]].

**Acknowledgments:** We thank Uwe Thumm, Sebastian Meuren and Andreas Kaldun for fruitful discussions. TM is grateful for the financial support of JSPS Fellowship for Research Abroad. GS acknowledges financial support by the Alexander von Humboldt Foundation (Project "Tirinto").

# Supplementary Materials for: "Molecular wave-packet dynamics on laser-controlled transition states"

**Methods:** In the experiment we use sub-*10* fs infrared (IR) pulses for high-order harmonic generation [1–5] in argon. The resulting attosecond pulse trains (APTs) have extreme ultra-violet (XUV) photon energies between *16* and *40* eV and are emitted in every half-cycle of the IR driving laser-field. The XUV beam is then focused in the center of a reaction microscope intersecting a supersonic jet of hydrogen gas. The experimental setup is illustrated in Fig. S1.

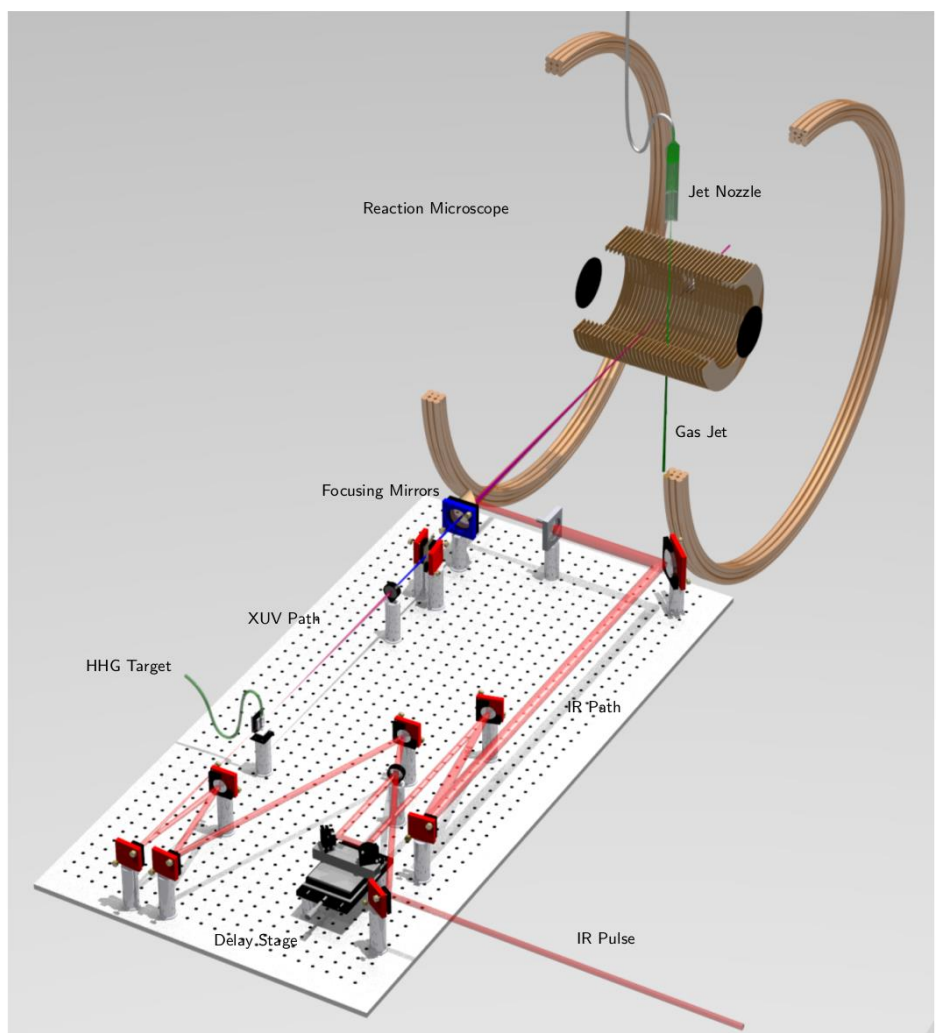

**Figure S1: Illustration of the experimental setup.** A near-infrared (IR) laser pulse is split into two using a broadband beam splitter. One copy of the laser pulse is used to create the attosecond pulse trains (APTs) in the high-order harmonic generation process in an argon HHG-target. Subsequently the APTs are focused, in the center of the reaction microscope, into a supersonic jet of molecular hydrogen. The other copy of the IR-pulse is time-delayed, and collinearly to the APTs, focused into the reaction microscope.

The hydrogen molecules are ionized by the APTs (pump pulse) launching the nuclear dynamics in the ground state of the molecular hydrogen ion. Subsequently, the molecular ion is dissociated by applying the fundamental IR pulse (control pulse) at a variable time delay τ after the pump pulse. Because the APTs are generated by the same IR pulse that is used to probe, the two laser pulses are intrinsically phase locked. Using a reaction microscope we obtain the full kinematic information of the process under investigation. In order to prepare the nuclear wave-packet dynamics in $H_2^+$ we consider the following reaction

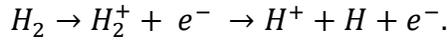
$$H_2 \rightarrow H_2^+ + e^- \rightarrow H^+ + H + e^-.$$

In the ionization, many vibrational levels of the $H_2(X\ ^1\Sigma_g^+)$ state are excited, according to the Franck-Condon overlap with the ground state nuclear wave function in the $H_2(X\ ^1\Sigma_g^+)$ potential. The IR pulse (polarized parallel to the XUV pulse) couples the $H_2^+(X\ ^2\Sigma_g^+)$ and the $H_2^+(A\ ^2\Sigma_u^+)$ state and causes the molecule to dissociate. This pump-probe scheme thus enables us to monitor the time evolution of the vibrational wave packet by varying the time delay τ between the two laser-pulses.

**Post-selection of laser intensity:** In the following, we demonstrate that we are able to study the influence of the IR-pulse intensity by selecting events in which the molecular axis is oriented at a specific angle θ to the polarization of the probe laser field.

Due to the dipole selection rules between the two states only parallel transitions, in which the molecular axis is oriented parallel to the effective electric field of the coupling laser, are allowed and, hence, the states |g> and the |u> are coupled solely by the electric field component parallel to the molecular axis. Thus, by selecting events in which the proton was detected under an angle θ with respect to the laser polarization, the effective laser intensity is reduced to $I_{eff} = I_0\ cos^2(\theta)$. Figure S2 shows the resulting Fourier spectra for three angles, i.e., increasing effective intensities from top to bottom, where the alignment was chosen to be in the intervals *0.2<cos(θ)<0.4*, *0.65<cos(θ)<0.75* and *cos(θ)>0.9*, in panels (a+b), (c+d), and (e+f), respectively. Excellent agreement between experimental data and the results of a quantum wave-packet dynamics simulation is found.

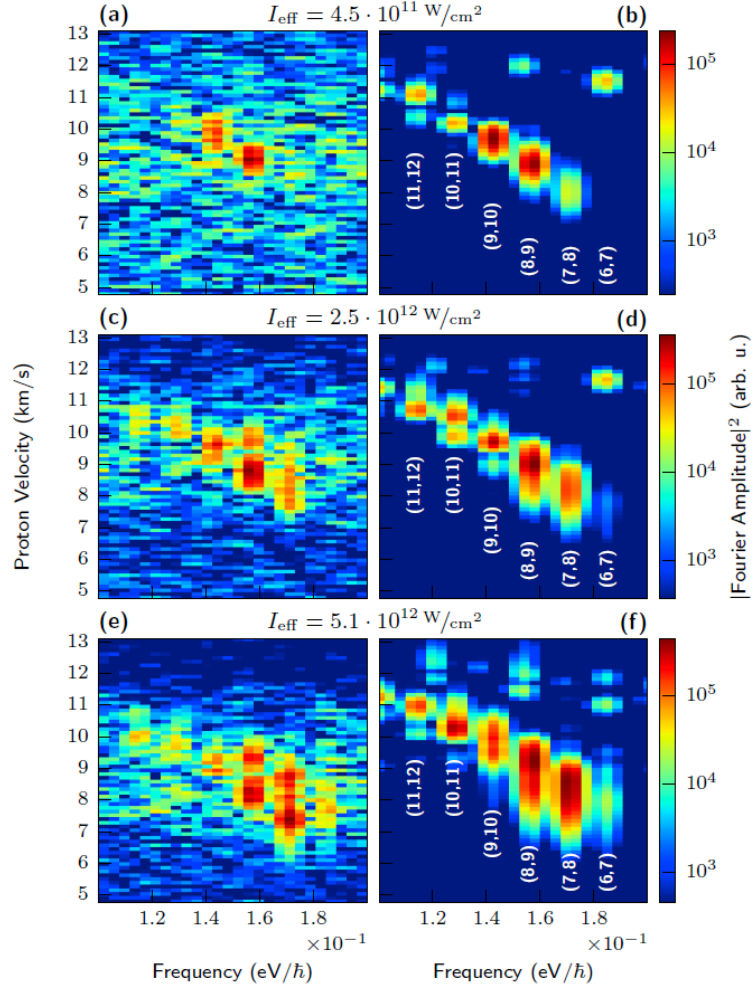

**Fig. S2: Fourier transforms of the time-dependent count distribution for different effective laser intensities.** The laser intensity increases from top to bottom. Left column: the obtained experimental data. Right column: the theoretical calculation using the TDSE is shown. With increasing field strength lower vibrational states get involved in the dissociation process. The pulse duration was $\sigma_{FWHM}$= 8.6fs with a center wavelength of $\lambda = 760nm$.

**Isolation of two-state dynamics:** Here, we demonstrate that the signal obtained by our Fourier filtering technique is indeed equivalent to the signal one would obtain by preparing and probing a single pair of vibrational states. For this, we compare the outcomes of wave packet dynamics simulations for two different cases: First, we prepare a superposition of two vibrational states, which we propagate by solving the time-dependent Schrödinger equation using the quantum wave-packet simulation [details to the methodology can be found in Ref. [6]]. The resulting count distribution is shown in Fig. S3 (a). As expected the obtained distribution is perfectly periodic. For comparability to the wavelet analysis presented in the main text, we subtracted the average along the time axis from the signal (equal to omitting the zero-frequency component in the wavelet analysis) and obtain Fig. S3 (c). Second, we simulate the full wave-packet dynamics, taking into account all populated vibrational states according to their Franck-Condon overlap to the neutral ground state wave function and then apply the wavelet analysis technique described in the main text [see of Fig. S3 (b)]. The resulting filtered signal [Fig. S3 (d)] is in very good agreement with the beating signal of two states after subtraction of the DC component [Fig. S3 (c)].

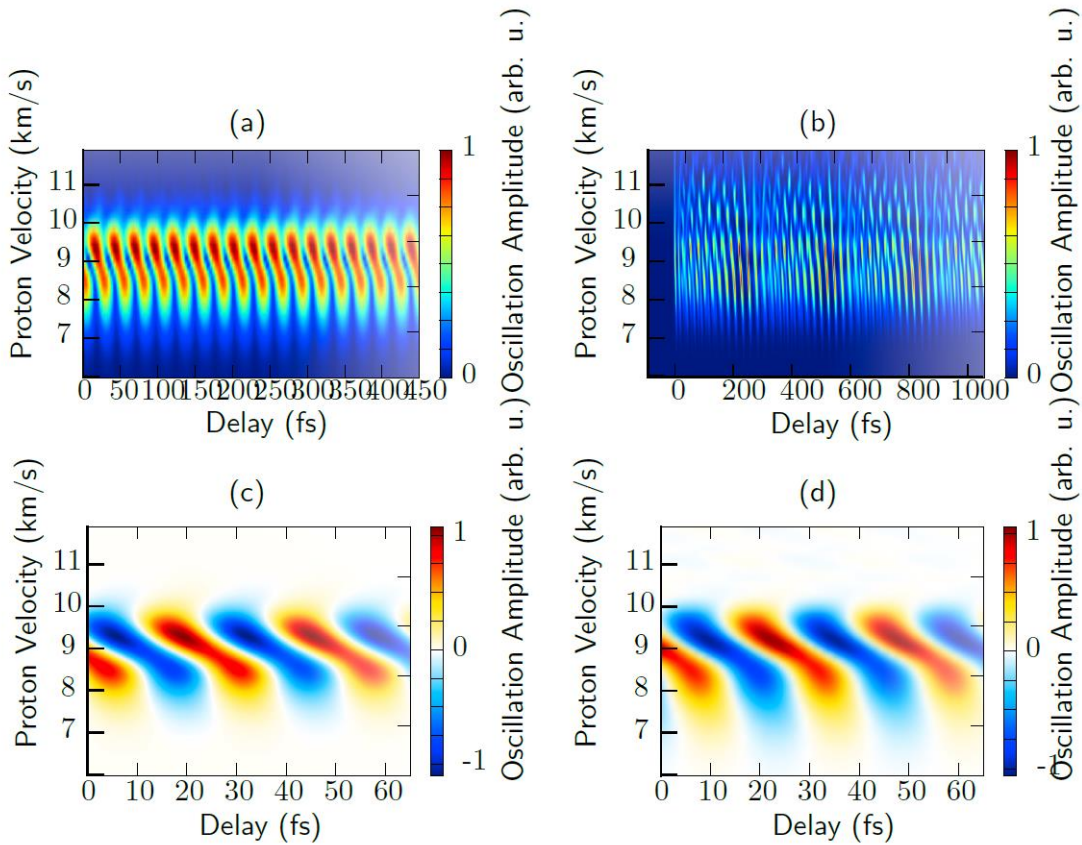

**Fig. S3: Comparison between the dynamics of two vibrational states and the Fourier filtered signal**. Left: Simulated count distribution resulting from the preparation of a wave packet consisting of two vibrational states ($v=8$ and $9$). (a) Bare spectrum and (c) spectrum after subtraction of the DC component. Right: (b) Count distribution for the full Franck-Condon wave-packet. (d) Result of the Fourier analysis. After the inverse transformation of the corresponding line in the Fourier spectrum to the time domain, the "two-state" signal is obtained.

**Illustration of Elevator Mechanism with Movies:**

**Mov. S4:** Movie S1 illustrates the elevator mechanism as acting in the semi-classical model for the case of a gain of total energy. The final velocity of the observed proton is the highest for this case.

**Mov. S5:** Illustration of the elevator for the case of a loss of total energy followed by a regain. The net change of the total energy vanishes.

**Mov. S6:** Finally the case of an energy loss resulting in the lowest final velocity of the observed proton is shown.